\documentclass[aps,prb,a4paper,10pt,twocolumn,showpacs,floatfix,bibnotes,superscriptaddress,preprintnumbers,longbibliography]{revtex4-2}
\usepackage{graphicx} 
\usepackage{dcolumn}  
\usepackage{bm}       
\usepackage{natbib}
\usepackage{soul}
\usepackage{color}
\usepackage[colorlinks,bookmarks=false,citecolor=darkblue,linkcolor=red,urlcolor=blue]{hyperref}

\usepackage[charter,greekuppercase=italicized]{mathdesign}

\usepackage{microtype}
\usepackage{afterpage}
\graphicspath{{../}{./}{figures/}}

\definecolor{darkred}{rgb}{0.7,0.0,0.0}

\definecolor{darkblue}{rgb}{0,0.02,0.45}

\definecolor{darkgreen}{rgb}{0.02,0.45,0.0}

\definecolor{violet}{rgb}{0.8,0.2,0.6}

\begin{document}

\title{Microscopic evidence of a field-induced critical spin-liquid state in a frustrated metal}

\author{I.\ Ishant}
\affiliation{\mbox{Department of Physics, Shiv Nadar Institution of Eminence, Gautam Buddha Nagar, Uttar Pradesh 201314, India}}

\author{Z.\ Guguchia}
\affiliation{PSI Center for Neutron and Muon Sciences CNM, 5232 Villigen PSI, Switzerland}

\author{V.\ Fritsch}
\affiliation{Experimental Physics VI, Center for Electronic Correlations and Magnetism, University of Augsburg, 86159 Augsburg, Germany}

\author{O.\ Stockert}
\affiliation{Max Planck Institute for Chemical Physics of Solids, 01187 Dresden, Germany}

\author{M.\ Majumder}
\email{mayukh.majumder@snu.edu.in}
\affiliation{\mbox{Department of Physics, Shiv Nadar Institution of Eminence, Gautam Buddha Nagar, Uttar Pradesh 201314, India}}

\date{\today}

\begin{abstract}
A field-induced quantum spin liquid (QSL) state is an extraordinary phenomenon, hitherto unobserved in metallic frustrated compounds. Recent bulk measurements have revealed intriguing field-induced magnetic states in metallic frustrated CePdAl. However, the nature of these field-induced states, potentially including a QSL state, remains unclear due to the lack of detailed microscopic investigation. To elucidate these field-induced states, we employed the transverse-field muon spin relaxation/rotation (TF-$\mu$SR) technique, applying various magnetic fields parallel to the c-axis in single-crystalline CePdAl over a broad temperature range (100~K-100~mK). Our $\mu$SR data indicate that field-induced low-temperature states for fields B$\leq B_{c2}(=3.4~T)$ exhibit long-range magnetic order, whereas for B>$B_{c2}$ they yield contrasting behavior. Notably, at 3.75 T, the transverse relaxation rate ($\lambda_T$) diverges following a power-law dependence below 800~mK along with an indication of finite frustration whereas the Knight shift is temperature independent. These observations corroborate the signature of a critical spin-liquid (CSL) with antiferromagnetic spin fluctuations. Furthermore, at 4.3 T, a non-Fermi liquid state is observed where frustration is absent. This comprehensive microscopic study strongly suggest the existence of a CSL state in a metallic frustrated system.

\end{abstract}

\maketitle

A quantum spin liquid (QSL) is an example of a highly entangled quantum state with exotic fractionalized excitations. The presence of frustration induced quantum fluctuations hinders any spontaneous symmetry breaking down to 0~K even though the spins are highly correlated~\cite{lancaster2023quantumspinliquids}. The experimental observation of such a state has been found in several insulating compounds with frustrated lattices (triangular, Kagome, pyrochlore, etc.) at ambient conditions~\cite{doi:10.1126/science.aay0668,PhysRevB.110.L060403}. It is quite rare that a static long-range magnetically ordered (LRO) state is suppressed and a possible QSL state is stabilized by the application of an external magnetic field. Such a field-tuning to a QSL state has been found e.g. in insulating $\alpha$-RuCl$_3$~\cite{PhysRevLett.119.037201,Banerjee2018} and Na$_3$Co$_2$SbO$_6$~\cite{PhysRevB.109.054411}. Interestingly, a QSL state which shows quantum critical behavior along with the presence of strong frustration without LRO~\cite{PhysRevLett.102.176401,CHUBUKOV1994601} and is named critical spin-liquid (CSL) state, has been observed in $\kappa$-(BEDT-TTF)$_2$Cu$_2$(CN)$_3$~\cite{Isono2016}. Even in some compounds with LRO (with milli Kelvin ordering temperature), the presence of a quantum-critical QSL state or proximate QSL state has been claimed to be observed just above the ordering temperature e.g. Cs$_2$CuCl$_4$~\cite{PhysRevB.68.134424}, K$_2$Ni$_2$(SO$_4$)$_3$~\cite{PhysRevLett.127.157204} and KYbSe$_2$~\cite{Scheie2024}. In general, the observation of a QSL or CSL is rarely seen in metallic frustrated compounds. At ambient conditions; CeRhSn~\cite{doi:10.1126/sciadv.1500001}, CeIrSn~\cite{PhysRevLett.126.217202} and Pr$_2$Ir$_2$O$_7$~\cite{Tokiwa2014} are the only examples. To the best of our knowledge, evidence for a field-driven QSL or CSL state has not been observed so far in any metallic frustrated compound. Thus, the stabilization of a field-induced QSL or CSL is an open question and of high importance. 

\begin{figure}
{\centering {\includegraphics[width=11cm]{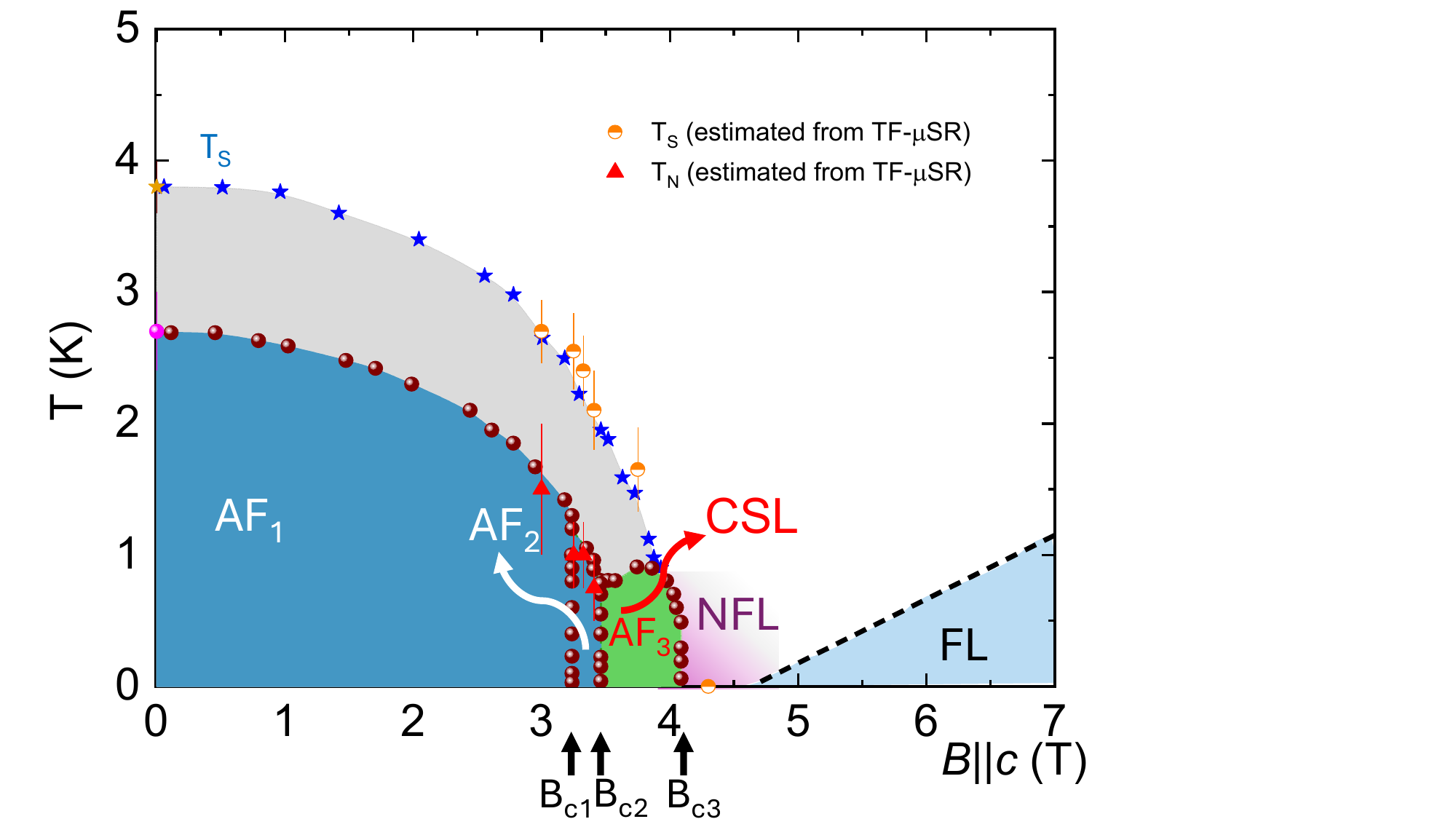}}\par} \caption{\label{fig:long_relax} Temperature-field magnetic phase diagram has been constructed from ref.~\cite{PhysRevLett.118.107204} for CePdAl when applied magnetic field is along the crystalligraphic c-axis. Blue stars and brown circles represent $T_s$ and $T_N$ from bulk measurements~\cite{PhysRevLett.118.107204}. Orange star and pink circle represent $T_s$ and $T_N$ estimated from zero field $\mu$SR measurements~\cite{PhysRevB.105.L180402}. NFL, FL and CSL represents non-Fermi liquid, Fermi-liquid and critical spin liquid respectively.}
\end{figure}

\begin{figure*}
{\centering {\includegraphics[width=15cm]{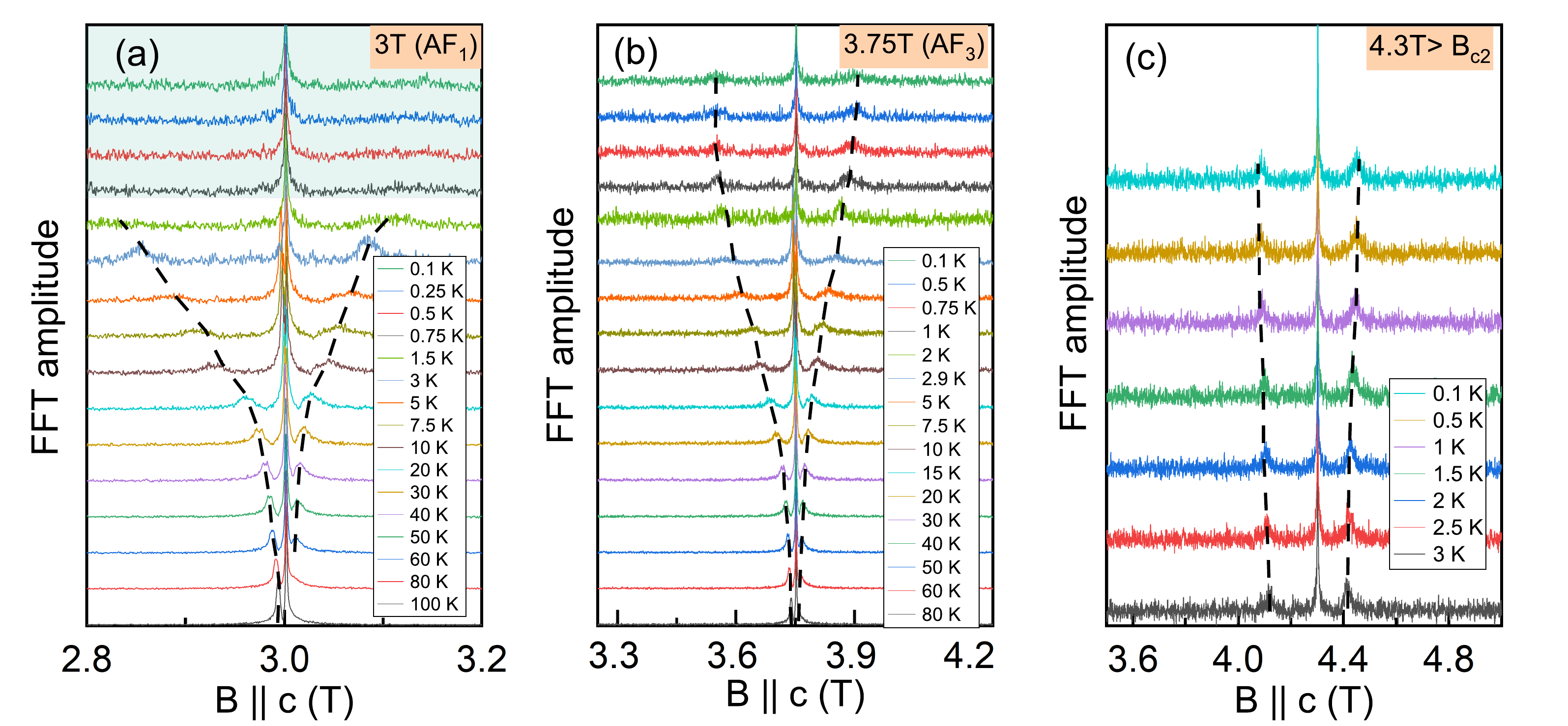}}\par} \caption{\label{fig:long_relax} (a), (b) and (c) show the temperature evolution of FFT $\mu$SR spectra at an applied field of 3~T, 3.75~T and 4.3~T respectively. The dashed lines are the guide to eyes and the shaded region in (a) represents LRO state. The left peak (line-1) and right peak (line-2) represent site-1 and site-2 respectively and the middle peak is the background contribution.}
\end{figure*}

In this respect, CePdAl is a promising candidate. The compound CePdAl, featuring Ce moments arranged in a highly frustrated kagome structure, presents a unique opportunity to investigate the interplay between magnetic frustration, RKKY,  Kondo interactions and magnetic anisotropy. With an antiferromagnetic transition temperature $T_\mathrm{N} \approx 2.7$\,K, CePdAl is exceptionally sensitive to tuning~\cite{KITAZAWA199428,Fritsch_2017,GOTO20021159}. Notably, hydrostatic and chemical pressure (Ni in place of Pd) suppresses LRO in CePdAl, giving rise to a QCP at a critical pressure ($P_c$) or concentration ($x_c$) of around 0.9 GPa and 0.15 ($x_c$) respectively~\cite{PhysRevB.105.L180402,PhysRevResearch.6.023112,Zhao2019}. Furthermore, an extensive non-Fermi liquid regime characterized by critical QSL-like dynamic correlations persists up to approximately twice the $P_c$ and concentration up to x=0.18. This unusual phase is attributed to the frustration inherent in the kagome lattice. Thus, by introducing frustration, the QSL state is stabilized, and the generic phase diagram of a quantum critical point (QCP) is drastically changed. Notably, frustration induces a NFL regime extending far beyond the critical point, defying conventional expectations. However, it is important to mention that doping inherently introduces disorder in the sample, whereas hydrostatic pressure applied to polycrystalline samples offers a cleaner tuning tool. But for a system exhibiting strong magnetic anisotropy, such as CePdAl, hydrostatic pressure is not an ideal tuning parameter for elucidating the nature of exotic states undoubtedly. To overcome these limitations of doping and hydrostatic pressure as a tuning parameter, we propose applying a magnetic field along a preferred direction of a single crystal, enabling precise identification of novel magnetic states. Specifically, our research aims to probe the field-tuned magnetic low-temperature state along the c-axis of single crystalline CePdAl. As illustrated in Fig. 1, bulk measurements have revealed a rich landscape comprising distinct antiferromagnetic (AF) phases ($AF_1$, $AF_2$, and $AF_3$) and hinting at the possible emergence of a QSL state around $B_{c2}$ where frustration is enhanced~\cite{PhysRevLett.118.107204}. Moreover, it was also pointed out that the Kondo scale gets suppressed by the application of an applied magnetic field at $B_K \approx 2.5$ T and thus above 2.5 T, the enhanced frustration (in the absence of Kondo screening) is responsible for the field-induced exotic states mentioned above~\cite{PhysRevLett.118.107204}. Notably, despite these intriguing observations, a detailed microscopic investigation to elucidate the nature of these low-temperature phases and to identify the presence of a QSL state in CePdAl has been conspicuously absent, motivating our present study.

To elucidate the intricate field-induced magnetic behavior of CePdAl, we employed the local probe muon spin relaxation/rotation ($\mu$SR) at the Swiss Muon Source (S$\mu$S), Paul Scherrer Institute, Switzerland. Transverse field (TF)-$\mu$SR measurements were performed, leveraging its unique capability to provide insightful information on local spin susceptibility through the estimation of Knight shift ($K_\mu$) and dynamical excitations via the transverse relaxation rate ($\lambda_T$). Utilizing the HAL-9500 spectrometer, we performed an extensive investigation on a high-quality single crystal of CePdAl, applying six carefully selected magnetic fields below and above $B_{c2}$: 3 T($< B_{c2}$), 3.25 T ($< B_{c2}$), 3.325 T ($< B_{c2}$), 3.4 T($= B_{c2}$), 3.75 T($> B_{c2}$), and 4.3 T($> B_{c2}$) with the magnetic field applied along the crystallographic c-axis. Note that the sample first was cooled down to 100~mK in the absence of the field and then the data have been taken while warming up after applying the field. This systematic study spanned over a wide range of temperatures between 100~mK and 100~K, allowing us to probe the field-temperature phase diagram and provide distinct answers to the outstanding questions regarding the spin liquid regime, non-Fermi liquid behavior, and magnetic phases. As summarized in Fig. 1, our extensive study indicates that all low-temperature states for B$\leq B_{c2}$ are LRO states (which correspond includes AF$_1$, AF$_2$ as shown in Fig. 1) whereas the states for B>$B_{c2}$ do not show any sign of LRO. Even more interestingly, the AF$_3$ state resembles the behaviour expected for a CSL state (hitherto unobserved in any metallic frustrated compound stabilized by applied magnetic field) and the state above the field B$_{c3}$ shows NFL behaviour without  frustration.  

The TF-$\mu$SR spectra were accurately modeled using a three-cosine functional form, as follows:

\begin{figure*}
{\centering {\includegraphics[width=15cm]{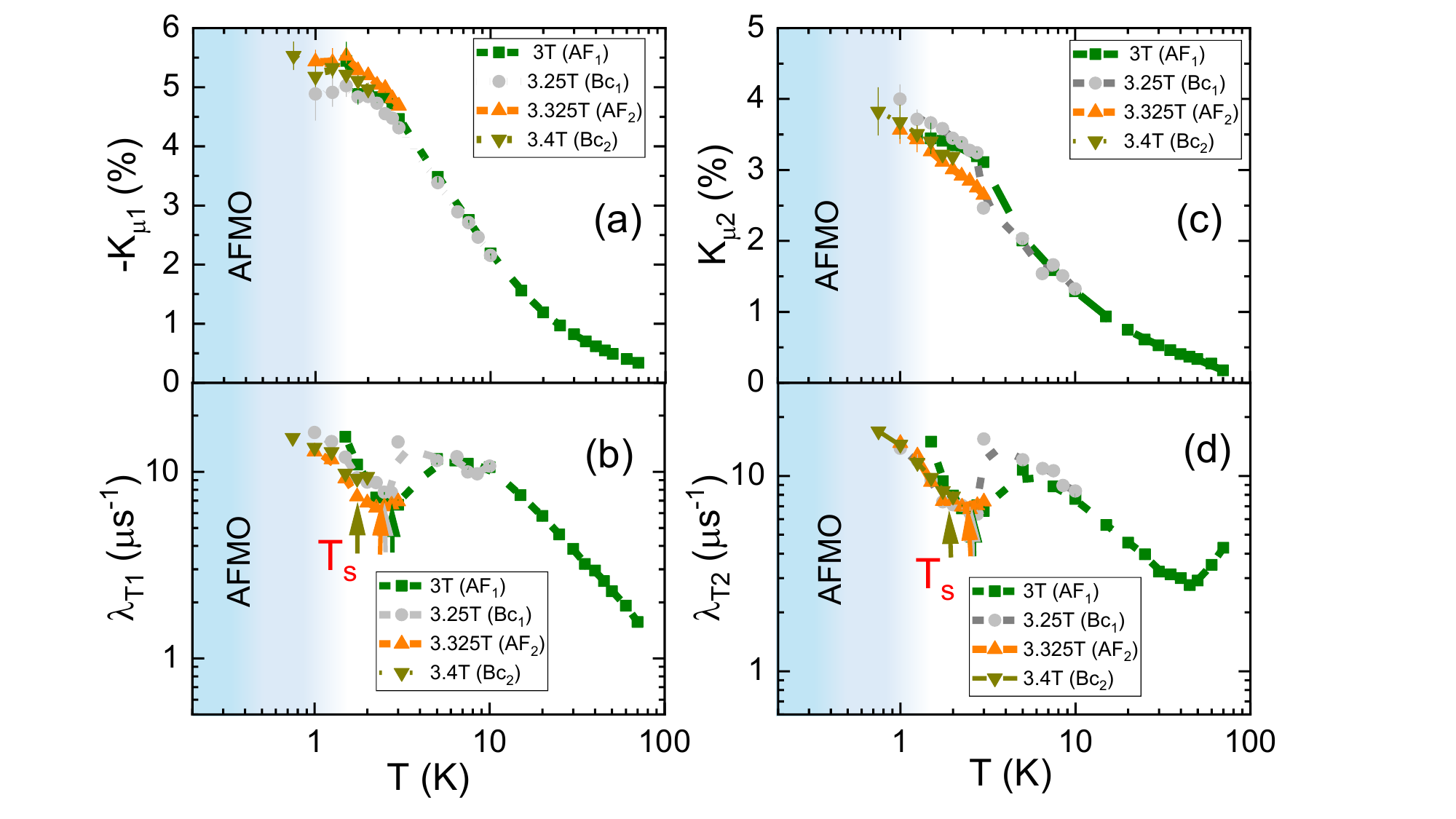}}\par} \caption{\label{fig:long_relax}(a) and (b) show the temperature dependence of knight shift ($K_\mu$\%) and relaxation rate ($\lambda_T$) at different magnetic fields (B$\leq B_{c2}$) for site-1. (c) and (d) represent the temperature dependence of knight shift (K\%) and relaxation rate ($\lambda_T$) at different magnetic fields (B$\leq B_{c2}$) for site-2. $T_s$ depicts a energy scale discussed in text. AFMO represent the antiferromagnetic ordered state. The dashed lines are the guide to eyes.}
\end{figure*}

\begin{eqnarray}
A(t) = & A [f_\mathrm{1}{\cos (2\pi\nu_1 t + \phi)e^{-\lambda_\mathrm{T1} t}} +  f_\mathrm{bkg}{\cos (2\pi\nu_{bkg} t + \phi)e^{-\lambda_\mathrm{bkg} t}} \nonumber \\
  &  + (1-f_\mathrm{1}-f_\mathrm{bkg}){\cos (2\pi\nu_2 t + \phi)e^{-\lambda_\mathrm{T2} t}}].  
\end{eqnarray}

Here, $A(t)$ is the time-dependent muon-spin polarization, with A the initial asymmetry. $f_\mathrm{1,bkg}$, $\nu_1, \nu_{bkg}, \nu_2$, $\phi$, $\lambda_\mathrm{T1, T2,{bkg}}$ are the fractions of individual sites, the muon Larmor frequencies, the initial phase (fixed to zero), and the transverse relaxation rates respectively for the 3 sites (2 stopping sites in CePdAl, one background site). The resulting asymmetry spectra, transformed using Fast Fourier Transform (FFT), are presented in Fig. 2, revealing three distinct peaks in the paramagnetic state (FFT of the experimental data and the fitting curve can be seen in Fig. S1 of the supplementary material~\cite{supply}). Notably, one peak (the middle line in Fig. 2) coincides with the applied magnetic field and remains temperature-independent, originating from muon stopped in the silver plate background, thereby serving as a reference point.
The remaining two peaks (site-1 and site-2) arise from muons occupying interstitial sites in CePdAl. Importantly, these sites exhibit a temperature dependence relative to the reference frequency, indicative of the development of local spin susceptibility as temperature decreases. The shift of these peaks as a function of temperature is defined as the Knight shift, $K_\mu(\%) = \frac{\nu - \nu_0}{\nu_0}\times 100$, where $\nu$ represents the resonance frequencies of site-1 and site-2 and $\nu_0$ represents the resonance frequency of the background signal. The Knight shift is proportional to the local susceptibility produced at the muon site (details are given in \cite{supply}). 

Now, we will delve into the field and temperature dependence of the spectral lines, with a focus on elucidating the nature of the field-induced low-temperature phases. Based on experimental observations (presented below), we will categorize the entire field range into two distinct regimes, one range of fields (B) where B$\leq B_{c2}$ and the other for B>$B_{c2}$. 

We begin by examining the TF-$\mu$SR results for B$\leq B_{c2}$. Notably, at 3 T ($AF_1$), just below the temperature at which the system enters to the $AF_1$-state from high temperatures, line-1 and line-2 disappear as shown in Fig. 2(a) due to the enormous broadening of line width indicating a very broad distribution of static local fields experienced by both the muon sites that obscures the signal. Similar observation has been seen for other fields (where B$\leq B_{c2}$) e.g. 3.25 T ($B_{c1}$), 3.325 T ($AF_2$), 3.4 T ($B_{c2}$). This behavior indicates that the low-temperature states for B$\leq B_{c2}$ are all static long-range magnetically ordered states. The temperature below which the lines become invisible can be considered as the ordering temperature ($T_N$ as shown in Fig. 1) which agrees well with the previous report from bulk measurements~\cite{PhysRevLett.118.107204}. We have plotted the Knight shift ($K_\mu$(\%)) as a function of temperature in Fig. 3(a,b), revealing that as the temperature decreases, the Knight shift for both sites exhibits a pronounced increase at an applied magnetic fields where B$\leq B_{c2}$. However, upon entering the $AF_1$ and $AF_2$ states, the lines unexpectedly vanish due to the broadening of the local field distribution as shown in Fig. 2(a) and 2(b) no longer allowing to determine $K_\mu$. The transverse relaxation rates, sensitive to probe low-energy spin fluctuations, exhibit a striking temperature dependence for different applied magnetic fields. With reducing temperature, the relaxation rates ($\lambda_{T1,T2}$) increase for both sites, as shown in Fig. 3(c,d), indicating the development of spin correlations. Notably, a pronounced downturn, or "inverted hump", emerges at intermediate temperatures, signaling the presence of short-range quasistatic ordering. Interestingly, this energy scale matches (as shown in Fig. 1) with the energy scale of the hump observed in the susceptibility measurements which was associated with the formation of spin-liquid-like correlations and indicates the presence of frustration~\cite{PhysRevLett.118.107204}. Thus, we also defined this energy scale as $T_{S}$ as in ref.~\cite{PhysRevLett.118.107204} and show the data in Fig. 1 (procedure to estimate $T_S$ has been mentioned in~\cite{supply}). Below $T_{S}$, $\lambda_{T1,T2}$ diverges at the long-range ordering temperatures due to the critical slowing down of spin fluctuations approaching the critical temperatures. This observation further corroborates the disappearance of both satellite lines and provide evidence of the long-range ordered state for all the fields in the range of B$\leq B_{c2}$.  

Moving on to the TF-$\mu$SR results for fields exceeding B>$B_{c2}$. In stark contrast to the behavior for fields B$\leq B_{c2}$, at higher fields (B>$B_{c2}$), namely 3.75 T ($AF_3$) and 4.3 T (>$B_{c3}$), line-1 and line-2 persist as shown in Fig. 1(b,c) down to the lowest temperature measured, suggesting the absence of the development of a broad distribution of local magnetic fields at the muon sites, thereby ruling out long-range magnetic ordering at low temperatures. This raises a fundamental question - \textit{what is the nature of these states above $B_{c2}$?}


\begin{figure*}
{\centering {\includegraphics[width=13cm]{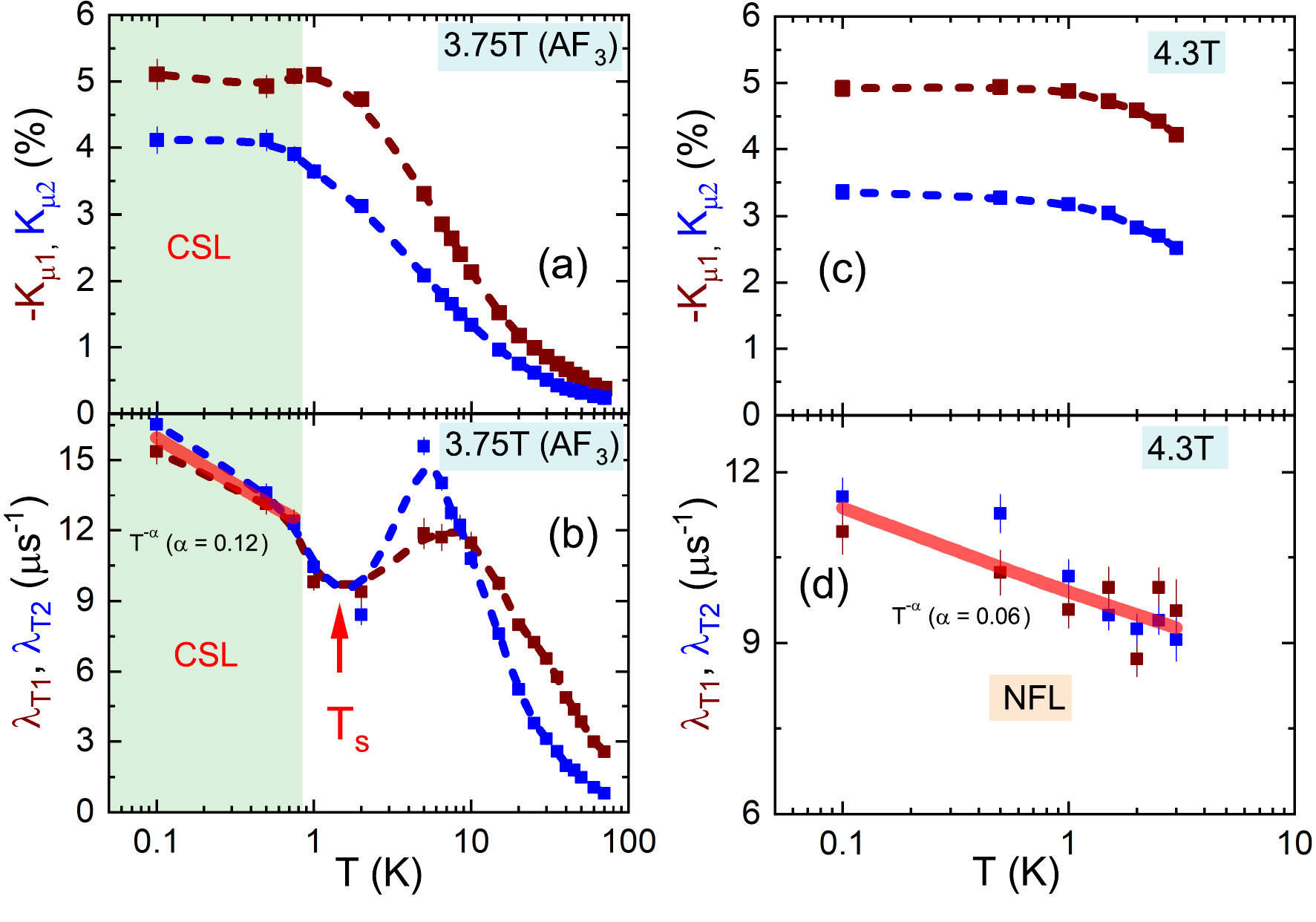}}\par} \caption{\label{fig:long_relax}(a) and (b) represent the temperature evolution of knight shift and relaxation rate ($\lambda_T$) for both of sites at 3.75~T ($AF_3$). (c) and (d) represent the temperature evolution of knight shift and relaxation rate ($\lambda_T$) for both of sites at 4.3~T ($AF_3$). The dashed lines are the guide to eyes. The red lines represents power-law behavior.}
\end{figure*}

We will first elucidate the nature of the $AF_3$ state - the persistence of frustration is reflected from the presence of $T_S$ in the $\lambda_{T1,T2}$ data (shown in Fig. 4(b)) along with the Knight shift which displays a remarkable plateau at low temperatures below 900~mK as shown in Fig. 4(a), unequivocally indicating the emergence of a spin-liquid regime with finite density of states of spinon Fermi surface\cite{PhysRevB.96.174411}. Even more interestingly, $\lambda_{T1,T2}$ shows a weak but finite divergence down to the lowest-temperature measured (100~mK) following a power-law ($\lambda_T \propto$ T$^{-\alpha}$) behavior (Fig. 4(c)). Such a critical power-law behavior in specific heat over temperature, $\mu$SR and NMR relaxation rates is expected to be seen in the CSL state as was predicted theoretically~\cite{PhysRevLett.102.176401} and observed experimentally~\cite{Pratt2011,PhysRevLett.126.217202,Isono2016,PhysRevB.106.064436}. Thus the observed critical divergence provides a compelling evidence that the QSL state ($AF_3$) is a critical quantum spin liquid (CSL) in contrast to a QSL away from the criticality, where one would expect to have a saturation of $\lambda_T$ along with the saturation of Knight shift. Considering $\lambda_T$ is dominated by spin-fluctuations (in the correlated non-ordered state), one can expect  

\begin{equation}
\lambda_T \approx 1/T_1 = T |A_c|^2 \lim_{\omega\to 0} \sum_q \chi''(q,\omega)  
\end{equation}

where $\chi''(q,\omega)$ is the imaginary part of the dynamical spin susceptibility summed over the first Brillouin zone. Thus, $\lambda_T$ serves as a probe for spin-fluctuations across both $q=0$ and $q\neq0$ momentum transfers, whereas the Knight shift exclusively probes $q=0$ correlations. The saturation of $K_\mu$ and the concomitant enhancement or divergence of $\lambda_{T1,T2}$ indicate that critical spin-fluctuations in the $AF_3$ state are predominantly driven by antiferromagnetic ($q\neq0$) contributions. 

It should be noted that clear evidence of magnetic ordering has been observed from neutron diffraction measurements~\cite{PROKES2006359} in the AF3 state, and the nature of this ordering remains unchanged within the field range of 3.2~T to 3.8~T. In contrast, $\mu$SR measurements reveal distinct behavior in the same field range (as discussed above). This discrepancy between the two techniques suggests that the AF3 state (at 3.75~T) represents a unique phase where the magnetic order appears static on the neutron time scale but dynamic on the $\mu$SR time scale. Such contrasting behavior is not unexpected and has been reported in several systems with dynamical ground states~\cite{PhysRevB.100.184415,PhysRevB.109.134431,PhysRevLett.96.127202}, owing to the different energy and time scales probed by neutron scattering and $\mu$SR.


Notably, recent studies~\cite{PhysRevLett.118.107204,PhysRevB.109.054405} defined the frustration parameter in metallic systems as $T_S/T_N$ where $T_N$ represented the long-range ordering temperature. According to our observation, the AF$_3$ state is not a LRO state and thus the frustration parameter diverges in the AF$_3$ state (as the upper bound of $T_N$ can be considered as 100~mK which is the lowest studied temperature) and provides further evidence for the stabilization of a CSL due to strong frustration. Moreover, theoretical calculations predict a saturation of $\chi$ and a divergence of $C/T$, upon cooling, attributed to gapless spinon excitations~\cite{PhysRevB.104.165120}. These findings lend additional credence to our experimental results. Remarkably, our TF-$\mu$SR findings, supported by calculations based on the frustrated square lattice model, unequivocally establish the existence of a spin liquid phase~\cite{PhysRevB.109.014103}. This result significantly enhances our comprehension of this enigmatic regime. Notably, this phenomenon bears striking resemblance to observations in chemical and hydrostatic pressure studies~\cite{PhysRevB.105.L180402, PhysRevResearch.6.023112}, where critical spin liquid (CSL) behavior emerges above the quantum critical point.

Next, we investigate the state above $B_{c3}$ (specifically 4.3~T). Initially, we note that applying such a magnetic field reduces frustration, and consequently, the frustration parameter $T_S$ diminishes to zero at $B_{c3}$, as inferred from its field dependent trend as shown in Fig. 1. Consistent with this expectation, the $\lambda_{T1,T2}$ at 4.3~T exhibits no "inverted-hump" like signature associated with $T_S$ as depicted in Fig. 4(d) and thus indicates a complete absence of frustration at 4.3~T. This observation underscores the absence of frustration in forming field-induced states above $B_{c3}$. Notably, $K_\mu$ saturates below about 1~K whereas, $\lambda_{T1,T2}$ shows a weak yet distinct divergence upon cooling, signaling non-Fermi liquid (NFL) behaviour. Moreover, the enhancement of $\lambda_{T1,T2}$ and saturation of $K_\mu$ with reducing temperature reveal the dominance of antiferromagnetic-type spin-fluctuations in this NFL state at 4.3~T. Note that ac-susceptibility measurements at 4.3~T also observe a saturation which could indicate a signature of a QSL as proposed in ref.~\cite{PhysRevB.97.235117} or a Fermi liquid (FL) state. The possibility of FL state can be discarded as we have observed the divergence of $\lambda_T$ in this regime. Moreover, the possibility of a QSL state can also be dismissed, given that our data demonstrate the disappearance of the frustration at ($B_{c3}$), a field lower than 4.3~T. Therefore, we conclusively classify the 4.3~T phase as a NFL state, without frustration. The suppression of frustration, above 4~T, was also pointed out by neutron diffraction~\cite{PROKES2006359} and further supports the present observation. Close proximity to a critical field $B_{c3}$ could be associated with the NFL behaviour at 4.3~T. 

In conclusion, we employed the TF-$\mu$SR technique as a microscopic tool to elucidate the field-induced states by applying several different magnetic field parallel to c-axis in a single crystalline CePdAl over a wide range of temperatures (100~K-100~mK). The $\mu$SR spectra indicate the presence of two muon stopping sites in CePdAl. The temperature-dependent shift and the transverse relaxation rate ($\lambda_T$) have been estimated at different applied fields. Our findings categorize the field-induced low-temperature states into three different regimes- (i) below a field B$\leq B_{c2}(=3.4~T)$: resembling a static long-range magnetically ordered state with substantial local field distributions, (ii) at a field of 3.75~T close to $B_{c2}$: evidence of a critical spin liquid with antiferromagnetic spin fluctuations is established and (iii) at a field of 4.3~T (>$B_{c2}$): the non-Fermi-liquid state has been observed without the presence of frustration. Thus our extensive microscopic study provided an unambiguous evidence for the existence of a CSL state which is a rare occurrence in a metallic frustrated compound. Hence, our study and observation of the field-induced CSL state will stimulate theoretical investigations of the kagome lattice, focusing on the competing interactions of similar energy scale. Furthermore, our findings are expected to encourage more experimental research into emerging “frustrated metallic systems,” a largely unexplored area, compared to its insulating counterpart, that may host novel states of matter such as QSL, CSL, QCP, and complex magnetic orderings.

Acknowledgment: II and MM would like to acknowledge fruitful discussions with Khokan Bhattacharya and Pietro Bonf\'a regarding calculations related to dipolar and contact fields. The authors also gratefully acknowledge Robert Scheuermann for valuable discussions and assistance with $\mu$SR experiments.

\bibliography{TF_CePdAl_v3_references.bib}

\end{document}